\documentclass[aps,prl,a4paper,10pt,twocolumn,showpacs,floatfix,longbibliography,superscriptaddress,amsmath,amsfonts,amssymb,preprintnumbers]{revtex4-1}
\usepackage{float}

\usepackage{mathtools}

\DeclarePairedDelimiter\ket{\lvert}{\rangle}
\DeclarePairedDelimiterX\braket[2]{\langle}{\rangle}{#1 \delimsize\vert #2}

\usepackage{bm}
\usepackage{dcolumn,graphicx,color,booktabs,microtype,afterpage}

\usepackage{sidecap}
\usepackage[dvipsnames]{xcolor}

\renewcommand{\tablename}{Table}
\makeatletter\renewcommand{\fnum@figure}[1]{\figurename~\thefigure.~}\makeatother
\makeatletter\renewcommand{\fnum@table}[1]{\tablename~\thetable.}\makeatother

\newcount\hh \newcount\mm
\hh=\time \divide\hh by 60
\mm=\hh \multiply\mm by 60 \mm=-\mm
\advance\mm by \time
\def\now{\number\hh:\ifnum\mm<10{}0\fi\number\mm}

\usepackage[colorlinks,plainpages=false,linkcolor=blue,urlcolor=blue,citecolor=blue,pdfpagemode=UseNone,pdfstartview=FitBH]{hyperref}

\newcommand{\tcr}[1]{\textcolor{black}{#1}}

\newlength{\myfigwidth}
\begin{document}

\title{Time-reversal symmetry breaking in the new noncentrosymmetric superconductor Zr$_3$Ir}  
\author{T.\ Shang}\email[Corresponding authors:\\]{tian.shang@psi.ch}
\affiliation{Laboratory for Multiscale Materials Experiments, Paul Scherrer Institut, Villigen CH-5232, Switzerland}
%
\author{S.\ K.\ Ghosh}\email[Corresponding authors:\\]{S.Ghosh@kent.ac.uk}
\affiliation{School of Physical Sciences, University of Kent, Canterbury CT2 7NH, United Kingdom}
\author{J.\ Z.\ Zhao}
\affiliation{Co-Innovation Center for New Energetic Materials, Southwest University of Science and Technology, Mianyang, 621010, People's Republic of China} 

\author{L.-J.\ Chang}
\affiliation{Department of Physics, National Cheng Kung University, Tainan 70101, Taiwan}
\author{C.\ Baines}
\affiliation{Laboratory for Muon-Spin Spectroscopy, Paul Scherrer Institut, CH-5232 Villigen PSI, Switzerland}
\author{M.\ K.\ Lee}
\affiliation{Department of Physics, National Cheng Kung University, Tainan 70101, Taiwan}
\author{D.\ J.~Gawryluk}
\affiliation{Laboratory for Multiscale Materials Experiments, Paul Scherrer Institut, Villigen CH-5232, Switzerland}
\author{M.\ Shi}
\affiliation{Swiss Light Source, Paul Scherrer Institut, Villigen CH-5232, Switzerland}
\author{M.\ Medarde}
\affiliation{Laboratory for Multiscale Materials Experiments, Paul Scherrer Institut, Villigen CH-5232, Switzerland}
%
\author{J. Quintanilla}
\affiliation{School of Physical Sciences, University of Kent, Canterbury CT2 7NH, United Kingdom}
\author{T.\ Shiroka}
\affiliation{Laboratorium f\"ur Festk\"orperphysik, ETH Z\"urich, CH-8093 Zurich, Switzerland}
\affiliation{Paul Scherrer Institut, CH-5232 Villigen PSI, Switzerland}

\begin{abstract}
We report the discovery of Zr$_3$Ir as a new type of unconventional 
noncentrosymmetric superconductor (with $T_c = 2.3$\,K), here
investigated mostly via 
muon-spin rotation/relaxation ($\mu$SR) techniques. Its superconductivity 
was characterized using magnetic susceptibility, electrical resistivity, and 
heat capacity measurements. The low-temperature superfluid density, determined 
via transverse-field $\mu$SR and electronic specific heat, suggests a 
fully-gapped superconducting state. The spontaneous magnetic fields, 
revealed by zero-field $\mu$SR below $T_c$, indicate the breaking of time-reversal 
symmetry in Zr$_3$Ir and, hence, the unconventional nature of its superconductivity. 
By using symmetry arguments and electronic-structure calculations 
we obtain a superconducting order parameter that is fully
compatible with the experimental observations.
Hence, our results clearly suggest that Zr$_3$Ir represents a new member of 
noncentrosymmetric superconductors with broken time-reversal symmetry.   
\end{abstract}

%
%


\maketitle\enlargethispage{3pt}

\vspace{-5pt}

Unconventional superconductors, in addition to $U(1)$ gauge 
symmetry, also break other types of symmetry~\cite{Sigrist1991,Tsuei2000}.
Among them, the breaking of time-reversal symmetry (TRS) below $T_{c}$ 
has been widely studied, in particular by means of zero-field muon-spin relaxation 
(ZF-$\mu$SR). As a very sensitive technique, $\mu$SR is able to detect the tiny 
spontaneous magnetic fields appearing below the onset of superconductivity (SC). 
Unconventional superconductors known to exhibit 
TRS breaking include, e.g., Sr$_2$RuO$_4$~\cite{Luke1998}, 
PrOs$_4$Sb$_4$~\cite{aoki2003}, UPt$_3$~\cite{Luke1993}, LaNiGa$_2$~\cite{Hillier2012}, 
LaNiC$_2$, La$_7T_3$, and Re$T$ ($T$ = transition metal)~\cite{Hillier2009,Barker2015,Shang2018,ShangReNb,Singh2014,Singh2017,Singh2018Re6Ti,singh2018La7Rh3}. 
The latter three also represent typical examples of noncentrosymmetric 
superconductors (NCSCs). 
In this case, the lack of space-inversion symmetry 
leads to an electric 
field gradient and, hence, to an antisymmetric spin-orbit coupling (ASOC), 
which splits the Fermi surface with opposite 
spin configurations. Often 
the strength of ASOC exceeds the superconducting 
energy gap, and 
the pairing of electrons belonging to different 
spin-split bands results in a mixture of singlet and triplet states. Due to 
such mixed pairing, NCSCs can exhibit significantly different properties from
their conventional counterparts, e.g. a nodal superconducting  
gap~\cite{yuan2006,nishiyama2007,bonalde2005CePt3Si,K2Cr3As3Pen,K2Cr3As3MuSR}, 
upper critical fields exceeding the Pauli 
limit~\cite{bauer2004,Carnicom2018,Shang2018} or, as recently proposed, 
topological superconductivity~\cite{Ali2014,Sun2015,Kim2018}. 
\tcr{In turn, the structure and/or symmetry may be important in determining 
the effects of ASOC on the superconducting properties~\cite{Anand2014}.} 

In general, the breaking of time-reversal and spatial-inversion symmetries 
are not necessarily correlated. Indeed, many NCSCs,  
such as Mo$_3$Al$_2$C~\cite{bauer2010}, Mo$_3$Rh$_2$N~\cite{Shang2018b}, La$T$Si$_3$~\cite{Anand2011,Anand2014,Smidman2014}, and Mg$_{10}$Ir$_{19}$B$_{16}$~\cite{Acze2010} 
do not exhibit spontaneous magnetic fields in the superconducting state and hence 
TRS is preserved. \tcr{TRS breaking in NCSCs is supposed 
to arise mostly from unconventional pairing mechanisms. For example, \tcr{LaNiC$_2$ 
is proposed to be a pure nonunitary triplet SC \cite{Hillier2009,Quintanilla2010} with pairing between same spins in two different orbitals \cite{Weng2016}}}.


Despite numerous examples of NCSCs, to date only a few of them are known 
to break TRS in their superconducting state. The causes of such 
selectivity remain largely unknown. Therefore, the availability 
of a new NCSC with broken TRS, such as Zr$_3$Ir reported here, 
would improve our understanding of the interplay between the
different type{s
of symmetry. 
In this Letter, we report systematic studies of Zr$_3$Ir by means of magnetization, 
transport, thermodynamic and muon-spin relaxation ($\mu$SR) measurements. 
The key observation of spontaneous magnetic fields, revealed by zero-field 
(ZF) $\mu$SR, indicates that Zr$_3$Ir represents a \emph{new member} 
of the NCSC family, making it a benchmark for the 
current theories of 
TRS breaking and unconventional SC in NCSCs. 

%
Polycrystalline Zr$_{3}$Ir samples were prepared by arc melting method~\cite{Supple}. 
The crystal structure and the sample purity were checked via x-ray powder diffraction using a Bruker D8 diffractometer. 
Consistent with previous results~\cite{Cenzual1985}, Zr$_3$Ir 
crystallizes in a tetragonal $\alpha$-V$_3$S-type noncentrosymmetric 
structure with space group $I\bar{4}2m$ (121)~\cite{Supple}. 
The magnetic susceptibility, electrical resistivity, and specific heat 
measurements were performed on a Quantum Design magnetic and physical property measurement system. 
The $\mu$SR measurements were carried out on  the GPS and LTF spectrometers 
of the $\pi$M3 beam line at the Paul Scherrer Institut (PSI), Villigen, Switzerland.

%
%

\begin{figure}[tb]
  \centering
  \includegraphics[width=\myfigwidth]{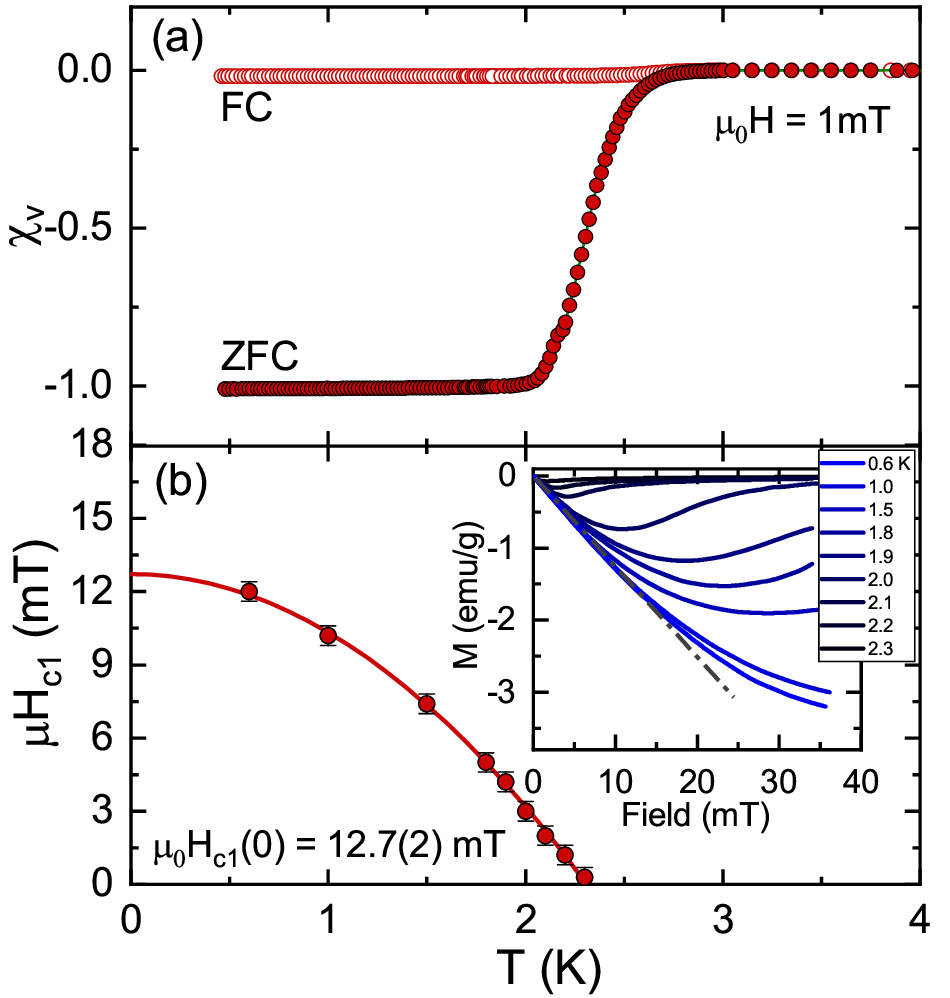}
  \caption{\label{fig:superconductivity}(a) Temperature-dependent 
  zero-field cooled (ZFC) and field-cooled (FC) magnetic susceptibilities, 
  measured in an applied field of 1\,mT for Zr$_3$Ir.  
  (b) Estimated $\mu_{0}H_{c1}$ values vs.\ temperature; the solid line 
  represents a fit to $\mu_{0}H_{c1}(T) =\mu_{0}H_{c1}(0)[1-(T/T_{c})^2]$. 
  The inset shows the field-dependent magnetization $M(H)$ recorded at 
  various temperatures. 
  $\mu_{0}H_{c1}$ was identified with the deviation of $M(H)$ from linearity (dashed-dotted line).}
\end{figure}
%

As shown in Fig.~\ref{fig:superconductivity}(a), the magnetic 
susceptibility indicates the SC onset at 2.7\,K, consistent 
with electrical resistivity data~\cite{Supple}. 
The splitting of FC- and ZFC susceptibilities is typical of type-II superconductors.
To determine $H_{c1}$, the field-dependent magnetization $M(H)$ was 
measured at various temperatures, 
as shown in 
the inset of Fig.~\ref{fig:superconductivity}(b). 
The solid line in Fig.~\ref{fig:superconductivity}(b) is a fit to 
$\mu_{0}H_{c1}(T) = \mu_{0}H_{c1}(0)[1-(T/T_{c})^2]$, which provides a 
lower critical field 12.7(1)\,mT and $T_c = 2.3$\,K. The bulk SC of Zr$_3$Ir was further 
confirmed by specific-heat measurements~\cite{Supple}.  

To explore the SC of Zr$_3$Ir at a microscopic
level, we resort to transverse field (TF) $\mu$SR measurements. 
Here, a FC-protocol is used to induce a flux-line lattice (FLL) in the mixed 
superconducting state. The optimal field value for such experiments was determined via preliminary field-dependent 
$\mu$SR measurements~\cite{Supple}. Figure~\ref{fig:TF_MuSR}(a) 
shows typical TF-$\mu$SR spectra, collected above and below $T_c$ 
at 30\,mT. The asymmetry of TF-$\mu$SR spectra is described by: 
\begin{equation}
\label{eq:TF}
A_\mathrm{TF} = A_\mathrm{s}  e^{- \sigma^2 t^2/2} \cos(\gamma_{\mu} B_\mathrm{s} t + \phi) +
A_\mathrm{bg} \cos(\gamma_{\mu} B_\mathrm{bg} t + \phi).
\end{equation}
Here $A_\mathrm{s}$ \tcr{(88\%)} and $A_\mathrm{bg}$ \tcr{(12\%)} are the sample and background  
asymmetries, with the latter not undergoing any depolarization. 
$\gamma_{\mu}/2\pi = 135.53$\,MHz/T is the muon gyromagnetic ratio, 
$B_\mathrm{s}$ and $B_\mathrm{bg}$ are the local fields sensed by 
implanted muons in the sample and the background (e.g., sample holder),
$\phi$ is a shared initial phase, and $\sigma$ is a Gaussian relaxation 
rate reflecting the field-distribution inside the sample. 

%
\begin{figure}[th]
	\centering
	\includegraphics[width=\myfigwidth]{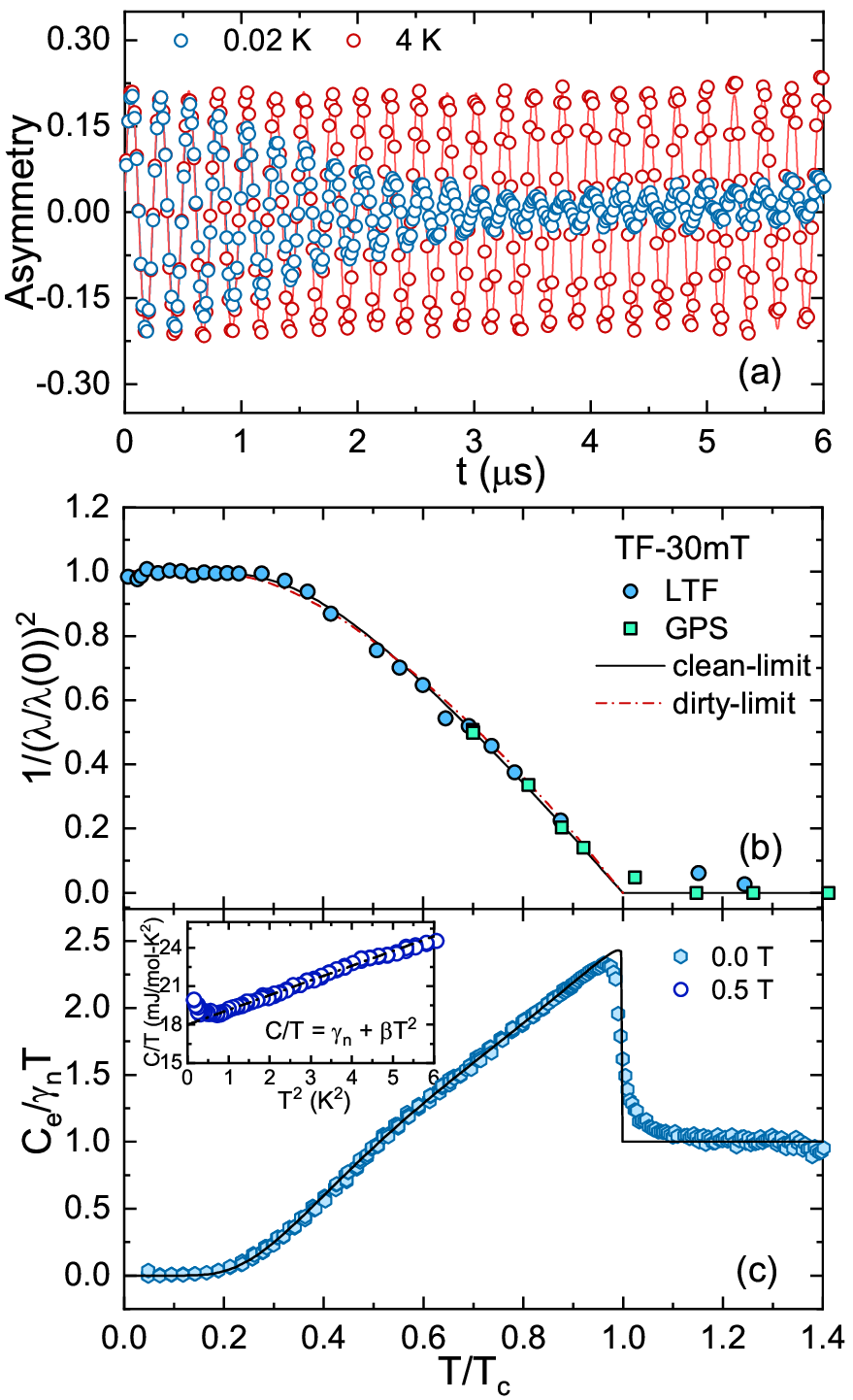}
	\vspace{-2ex}%
	\caption{\label{fig:TF_MuSR}(a) Time-domain TF-$\mu$SR spectra in the 
	superconducting (0.02\,K) and the normal (4\,K) phase of Zr$_{3}$Ir measured in a field of $H_\mathrm{appl} = 30$\,mT. 
	{Normalized} superfluid density (b) and zero-field electronic 
	specific heat (c) vs.\ the reduced temperature ($T$/$T_c$).  
	The $\mu$SR-datasets collected on GPS and LTF spectrometers are highly consistent. 
	The inset in (c) shows the raw $C/T$ data measured in a 0.5-T applied field as a function 
	of $T^{2}$. The dashed-line is a fit to $C/T = \gamma_{\mathrm{n}} + \beta T^{2}$, 
	from which the phonon contribution was evaluated. The solid black lines in (b) 
	and (c) represent fits using a fully-gapped $s$-wave model, while the red dash-dotted line in (b) is a fit to a dirty-limit model.}
\end{figure}
%

In the superconducting state,
$\sigma$ includes contributions from both the FLL ($\sigma_\mathrm{sc}$) and a 
smaller, temperature-independent relaxation, due to nuclear moments 
($\sigma_\mathrm{n}$). The former 
can be extracted by subtracting the nuclear contribution in quadrature, 
i.e., $\sigma_\mathrm{sc}$ = $\sqrt{\sigma^{2} - \sigma^{2}_\mathrm{n}}$. 
Since the upper critical field $\mu$$H_{c2}$ of Zr$_3$Ir is
relatively modest (0.62\,T)~\cite{Supple},   
to extract the magnetic penetration depth $\lambda_\mathrm{eff}$ from 
the measured $\sigma_\mathrm{sc}$ we had to consider the expression~\cite{Barford1988,Brandt2003}:
\begin{equation}
\label{eq:sig_to_lambda}
\sigma_\mathrm{sc} (h) = 0.172 \frac{\gamma_{\mu} \Phi_0}{2\pi}(1-h)[1+1.21(1-\sqrt{h})^3]\lambda^{-2}_\mathrm{eff}, 
\end{equation}
valid for intermediate values of the reduced magnetic field, 
$h = H_\mathrm{appl}/H_\mathrm{c2}$. 
Figure~\ref{fig:TF_MuSR}(b) shows the normalized superfluid density 
($\rho_{\mathrm{sc}} \propto 1/\lambda^{2}$) versus the reduced temperature 
$T/T_c$ for Zr$_3$Ir. 
For $T < 0.3\,T_c$, the superfluid density is nearly 
temperature independent, indicating the absence of  
residual low-$T$ excitations and, hence, a fully-gapped superconducting state. 
The tem\-per\-a\-ture\--de\-pen\-dent superfluid density was 
fitted by using a fully-gapped $s$-wave model with a single superconducting gap, which provides 
$\Delta(0) = 0.30(1)$\,meV and $\lambda(0) = 294(2)$\,nm.
\tcr{The coherence length $\xi_0$ is slightly larger than the electronic mean 
free path $l_\mathrm{e}$ (see Table SIII), implying that Zr$_3$Ir is  
in the dirty limit. Thus, similar to other NCSCs~\cite{Frandsen2015}, 
the temperature dependence of its superfluid density was also 
analyzed by a dirty-limit model, which yields a gap
value of 0.23(1) meV, slightly smaller than the clean-limit
value, but consistent with previous studies~\cite{Sajilesh2019}.
}


%
%

%
\begin{figure}[ht]
	\centering
	\includegraphics[width=\myfigwidth]{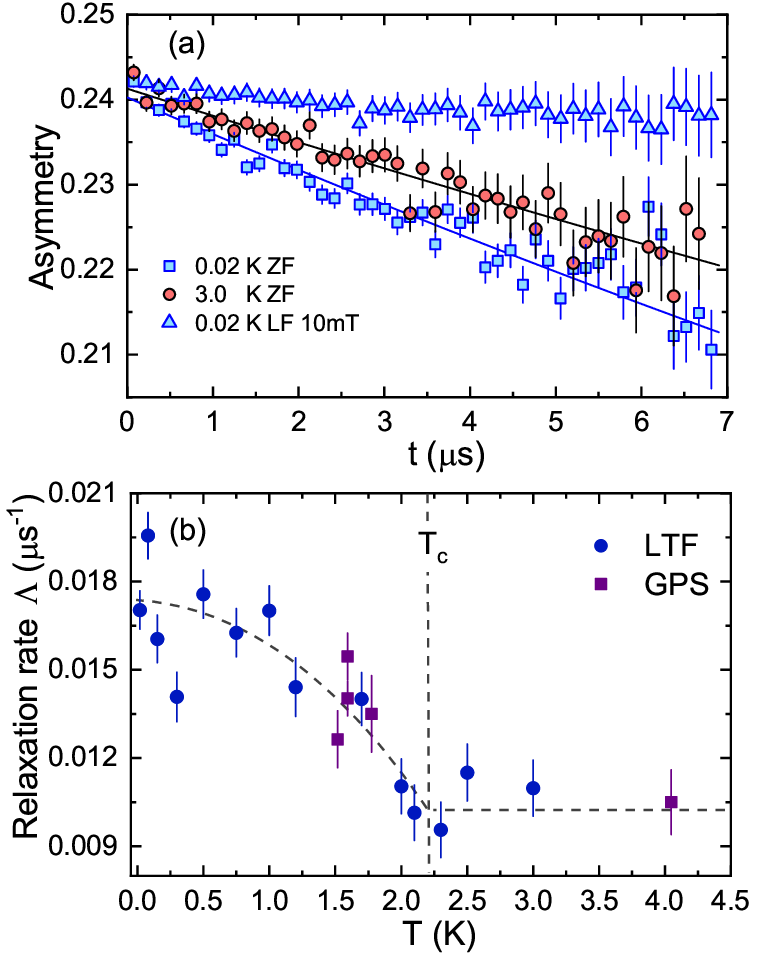}
	\vspace{-2ex}%
	\caption{\label{fig:ZF_muSR}Representative zero-field $\mu$SR spectra 
	in the superconducting (0.02\,K) and the normal (3\,K) phase of Zr$_3$Ir, 
	together with longitudinal field data, collected at 0.02\,K 
	and 10\,mT. The solid lines are fits to Eq.~(\ref{eq:Loren}). (b) Derived relaxation rate $\Lambda$ vs.\ 
	temperature. The dashed lines are guides to the eye. The datasets collected on GPS and LTF spectrometers are highly consistent.}
\end{figure}
%

\tcr{The zero-field specific-heat data after subtracting the phonon $\beta T^2$ contribution 
[see inset in Fig.~\ref{fig:TF_MuSR}(c)] are shown in Fig.~\ref{fig:TF_MuSR}(c). The fit (solid-line) yields 
a Sommerfeld coefficient} $\gamma_\mathrm{n} = 17.9$\,mJ\,mol$^{-1}$K$^{-2}$ and a single isotropic 
gap, $\Delta(0) = 0.32(1)$\,meV. Thus, both specific-heat and TF-$\mu$SR 
results are compatible with a \emph{fully-gapped} superconducting state.

%

ZF-$\mu$SR measurements were used to establish  
the onset of TRS breaking in the superconducting state 
through its key signature, the appearance of 
spontaneous magnetic fields below $T_{c}$. 
Representative ZF-$\mu$SR spectra, shown in Fig.~\ref{fig:ZF_muSR}(a),  
indicate a clear change in the muon-spin relaxation between 3\,K and 0.02\,K. 
For nonmagnetic materials, in the absence 
of applied fields, the depolarization of muon spins is mainly determined 
by the randomly oriented nuclear moments. This behavior is normally 
described by a Gaussian Kubo-Toyabe relaxation function~\cite{Kubo1967,Yaouanc2011}, 
as in the case of Re-based NCSCs~\cite{Singh2014,Singh2017,Singh2018Re6Ti,Shang2018,ShangReNb}. 
For Zr$_3$Ir, the depolarization shown in Fig.~\ref{fig:ZF_muSR} 
is more consistent 
with a Lorentzian decay.  
Indeed, attempts to analyze the data with  
a combined Gaussian and Lorentzian 
Kubo-Toyabe function, as in Refs.~\onlinecite{Shang2018,ShangReNb},  
systematically exclude the Gaussian component. This suggests that 
the fields sensed by the implanted muons arise from 
the diluted (and tiny) nuclear moments present in Zr$_3$Ir. Therefore, the solid lines 
in Fig.~\ref{fig:ZF_muSR}(a) are fits to a Lorentzian Kubo-Toyabe relaxation function:
\begin{equation}
\label{eq:Loren}
A_\mathrm{ZF} = A_\mathrm{s}\left\lbrace\frac{1}{3} + \frac{2}{3}\,(1 - \Lambda t)\,\mathrm{e}^{- \Lambda t}\right\rbrace + A_\mathrm{bg}.
\end{equation}
Here $A_\mathrm{s}$ and $A_\mathrm{bg}$ are the same as in the TF-$\mu$SR 
case (see Eq.~\ref{eq:TF}). 
The derived muon-spin 
relaxation rate $\Lambda$ vs. temperature is summarized in 
Fig.~\ref{fig:ZF_muSR}(b). Its relative change is comparable to that in other 
NCSCs with broken TRS~\cite{Barker2015, Singh2014, Singh2017,Singh2018Re6Ti,Shang2018,Hillier2009,ShangReNb,singh2018La7Rh3}, and  $\Lambda(T)$ shows a distinct increase below $T_c$, while being almost temperature independent above $T_c$. Such increase in the
muon-spin relaxation rate below $T_c$ was also found in other compounds, e.g., La$_7$Ir$_3$, LaNiC$_2$, and Sr$_2$RuO$_4$~\cite{Barker2015,Hillier2009,Luke1998}. This provides 
unambiguous evidence that TRS is also broken in the superconducting state 
of Zr$_3$Ir. 
\tcr{ We note that, due to insufficient time resolution, a previous 
study reported standard deviations in $\Lambda(T)$ 
of $\sim$0.01$\mu$s$^{-1}$~\cite{Sajilesh2019}. Consequently, it could not 
capture the small yet systematic increase in $\Lambda(T)$, 
less than $\sim$0.008$\mu$s$^{-1}$, shown in Fig.~\ref{fig:ZF_muSR}(b)~\cite{Supple}.
}
To rule out the possibility of extrinsic effects related to a  
defect/impurity- induced relaxation at low temperatures, we also carried 
out auxiliary longitudinal-field $\mu$SR measurements at base temperature 
(0.02\,K). As shown in Fig.~\ref{fig:ZF_muSR}(a), a small 10\,mT field is sufficient to lock the muon spins and, hence, to fully decouple them 
from the weak spontaneous magnetic fields. This further supports the intrinsic nature of TRS breaking in the superconducting state of Zr$_3$Ir.

\begin{figure}[!h]
	\centering
     \includegraphics[width=1.2\myfigwidth]{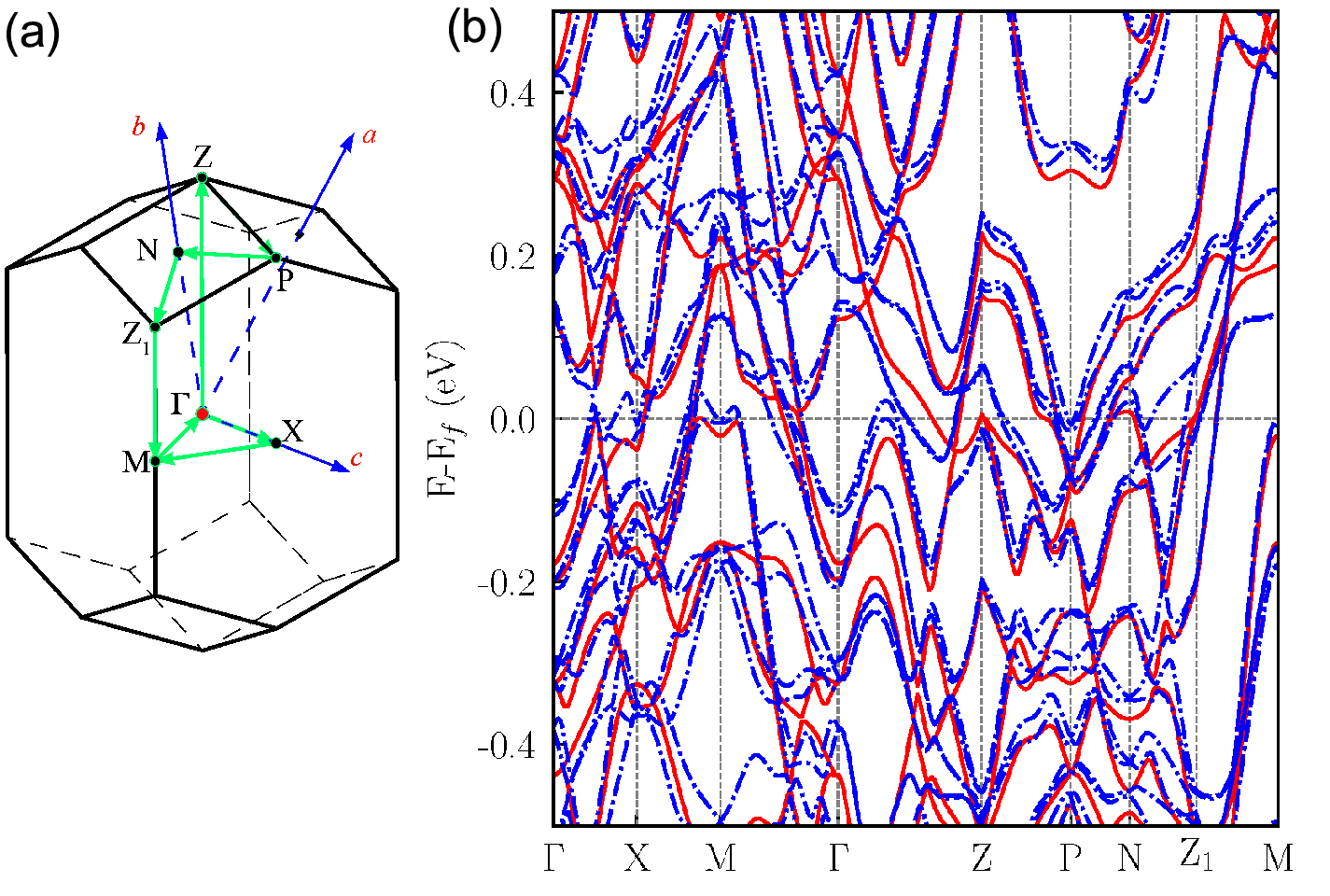}
	\vspace{-4ex}%
	\caption{\label{fig:bandstructure}(a) High-symmetry points of the Zr$_{3}$Ir unit cell. 
		(b) Electronic band structure with- (dotted blue lines) and without 
		SOC (solid red lines), within $\pm$0.5\,eV from the Fermi energy.}
\end{figure}
%

To date, TRS breaking has been encountered
only in three structurally 
different NCSCs: the CeNiC$_2$-type LaNiC$_2$~\cite{Hillier2009}, the $\alpha$-Mn-type Re$T$~\cite{Singh2014,Singh2017,Singh2018Re6Ti,Shang2018,ShangReNb}, and the Th$_7$Fe$_3$-type La$_7T_3$~\cite{Barker2015,singh2018La7Rh3}. With the $\alpha$-V$_3$S-type Zr$_3$Ir reported here, we show the TRS breaking to occur in a new \emph{structurally different} NCSC class. 
However, we point out that while crystal structure might influence 
TRS breaking in NCSCs, its role is not well known. 
Thus, many NCSCs, e.g., Mo$_3$Al$_2$C, Mo$_3$Rh$_2$N, La$T$Si$_3$, Mg$_{10}$Ir$_{19}$B$_{16}$, and Nb$_{0.5}$Os$_{0.5}$~\cite{bauer2010,Acze2010,Shang2018b,Anand2014,Smidman2014,SinghNbOs} 
preserve TRS although the last two share the same $\alpha$-Mn-type structure 
with other TRS-breaking Re$T$ NCSCs. \tcr{Moreover, the triplet pairing component appears to have a negligible effect in the properties of these NCSCs}. Finally, in 
some of NCSCs, such as Re$_3$W~\cite{ShangReNb,Biswas2012}, changes in muon-spin 
relaxation below $T_c$ may be below the resolution of the $\mu$SR technique.

To gain insight into the structure of the superconducting order 
parameter in Zr$_3$Ir, we performed electronic band-structure 
calculations and symmetry analysis of the Ginzburg-Landau (GL) free 
energy~\cite{Annett1990,Sigrist1991}.  
The band structures, both with and without SOC, were calculated by means 
of density-functional theory (DFT) within the generalized gradient approximation 
(GGA)~\cite{Supple,Perdew:1996iq,Kresse:1996kl,Kresse:1996vk,Kresse:1999wc,Blochl:1994zz} 
and are shown in Fig.~\ref{fig:bandstructure}.  
The estimated band splitting near the Fermi level due to the ASOC, which plays an important role in determining the superconducting properties, is about $100$\,meV. While it is comparable to that of NCSCs, such as PdBiSe ($\sim$109\,meV)~\cite{Kakihana2015} and K$_2$Cr$_3$As$_3$ ($\sim$ 60\,meV)~\cite{Jiang2015}, 
it is much smaller than that of CePt$_3$Si ($\sim$200\,meV)~\cite{Samokhin2004} and Li$_2$Pt$_3$B ($\sim 200$ meV)~\cite{Lee2005}.
The band structure shown in Fig.~\ref{fig:bandstructure}(b) also reveals multiple dispersive bands crossing the Fermi energy. In particular, the electron pockets centered around the $\Gamma$ point are much larger than the hole pockets centered around the $Z$ point.

\tcr{The space group of Zr$_3$Ir, $I\bar{4}2m$, is symmorphic (direct product 
of the corresponding point group $D_{2d}$ and the group of crystalline translations). Hence, the uniform superconducting instabilities of Zr$_3$Ir are
fully determined by the normal-state symmetry group $\mathcal{G} = G \otimes U(1) \otimes \mathcal{T}$, where $G$ is the group of point-symmetries and spin rotations, and $\mathcal{T}$ is the group of TRS. The GL free energy of the system must be invariant under $\mathcal{G}$, which has the same spatial symmetry as $D_{2d}$  having four 1D and one 2D irreducible representations (irreps). The presence of the 2D irrep allows for a two-component superconducting order parameter in Zr$_3$Ir, while a non-trivial phase difference between them 
may lead to
TRS breaking. Note that SC in this channel requires additional crystalline symmetry breaking and, hence, the pairing mechanism is necessarily \emph{unconventional} (i.e., not phonon-mediated)}.   

SOC has dramatic consequences on the pairing symmetry of Zr$_3$Ir 
due to noncentrosymmetry and its relatively large band splitting (see Fig.~\ref{fig:bandstructure}). When SOC cannot be neglected the superconducting state is, in general, a mixture of singlets and triplets.
The only possible order parameter breaking TRS in this case is 
$\hat{\Delta}(\bm{k}) = \left[ \Delta_0(\bm{k}) + \bm{d}(\bm{k}).\pmb{\sigma}\right] 
i \sigma_y$ where $\Delta_0 (\bm{k}) = A_1 (k_x + i k_y)k_z$ is the singlet component 
and $\bm{d}(\bm{k}) = \left[A_2 k_z, i A_2 k_z, B_2 (k_x + i k_y) \right]$ 
is the triplet component (the full derivation is given in the Supplemental Material). Here $\pmb{\sigma} = \{\sigma_x,\sigma_y,\sigma_z\}$ 
are the Pauli matrices and, $A_1$, $A_2$ and $B_2$ are constants independent 
of $\bm{k}$. \tcr{When $|A_2|,|B_2|\ll |A_1|$ the limit of weak SOC is recovered.}

Now, we discuss the compatibility of the above order parameter with the 
fully-gapped SC observed in Zr$_3$Ir. First, we consider
the cases of a singlet- or trip\-let\--do\-mi\-na\-ted order parameter.  A singlet-dominated TRS breaking order parameter ($A_2,B_2  \approx 0$) leads to an energy gap 
$|A_1||k_z| \sqrt{k^2_x + k^2_y}$. 
It has a line node at the ``equator'' for $k_z = 0$ and two point nodes at the ``north'' and ``south'' poles. Similarly, a triplet-dominated TRS breaking order 
parameter ($A_1 \approx 0$) corresponds  to an energy gap 
$[{g(k_x,k_y) + 2 A_2^2 k^2_z - 2 |A_2||k_z| \sqrt{g(k_x,k_y) + A_2^2 k^2_z}}]^{1/2}$, 
where $g(k_x,k_y) = B_2^2 (k^2_x + k^2_y)$. This instability also has 
two point nodes at the two poles, but no line nodes (see the 
respective gap plots in the Supplementary Material). 
In the general case, where both singlet- and triplet components are significant, we compute the excitation 
spectrum numerically, by using the Bogoliubov-de Gennes formalism~\cite{Sigrist1991}. 
We use the simplest form of normal-state band structure and 
ASOC-coupling constant compatible with the crystal symmetry~\cite{Supple}. 
As soon as a triplet component is present, the line node at the ``equator'' 
is gapped out. In contrast, the  ``north'' and ``south'' point nodes are present throughout the phase diagram.

\tcr{The above results are based only on
symmetry arguments considering a generic Fermi surface. 
To adapt them to Zr$_{3}$Ir, we consider its Fermi surfaces which 
contribute the most to the density of states at the Fermi level 
(computed using DFT~\cite{Supple}). They are \emph{open} at the two 
poles~\cite{Supple}, implying fully-gapped behavior is expected for TRS-breaking order parameters with a non-negligible 
triplet component. Note that an equatorial line node for a singlet-dominated SC order 
parameter, necessarily gives rise to a gapless spectrum for the 
given Fermi-surface topology}. Therefore, to reproduce the 
fully-gapped spectrum of Zr$_{3}$Ir,  
a triplet component (possibly small) induced by SOC is essential.

The specific-heat jump 
{at $T_c$ ($\Delta C/\gamma$T$_{c}$ $\sim$ 1.32) and the gap to 
critical-temperature ratio ($2\Delta/\mathrm{k}_\mathrm{B}T_{c}$ $\sim$  3.24) 
of Zr$_3$Ir} shown in Fig.~\ref{fig:TF_MuSR} seem to suggest a
singlet, phonon-mediated SC. This raises
the question of how such a conventional mechanism may lead to a state 
with broken TRS, corresponding to a non-trivial irrep of the crystal 
point group. Recently, it was proposed that in multi-band systems, whose 
bands derive from distinct but symmetry-related sites within the unit cell, 
this may be achieved by a loop super-current state~\cite{Ghosh2018}. 
Interestingly, such conditions are satisfied in Zr$_3$Ir. In this picture, the $k$-dependent order parameter comes from a real-space pairing potential:
$\ket{\Delta} = \ket{1}+i\ket{2}$ where $\ket{1} = (0,0,-1,1)$ and 
$\ket{2} = (-1,1,0,0)$ are real-space basis functions giving the on-site, 
singlet pairing strength in each of the four symmetry-related sites 
within the unit cell~\cite{Supple}. This ground state has finite 
currents within a unit cell spontaneously breaking TRS at $T_c$~\cite{Ghosh2018}. Note that the energy of this superconducting} instability is 
driven by singlet pairing, 
while the additional triplet contribution is induced by SOC. 

Finally, for weak SOC, $G = D_{2d} \otimes SO(3)$, with 
$D_{2d}$ and $SO(3)$ acting independently. Hence, $G$ can have
1, 2, 3, and 6-dimensional irreps. As a result, additional TRS-breaking SC states 
may appear, including those arising from a pure $SO(3)$\--pair\-ing, as proposed for LaNiC$_2$ and LaNiGa$_2$~\cite{Hillier2009,Quintanilla2010,Hillier2012}.

In conclusion, we have discovered a new structurally different member 
of the NCSC class, Zr$_3$Ir, which breaks time-reversal symmetry at the superconducting transition. The spontaneous magnetic fields 
appearing below $T_c$ were detected by ZF-$\mu$SR, while its electronic 
properties were investigated by means of magnetization, transport, 
thermodynamic, and $\mu$SR measurements. Both the zero-field specific-heat 
and superfluid density (from TF-$\mu$SR) reveal a single-, fully-gapped
superconducting state in Zr$_3$Ir. 
Theoretically, we obtain a superconducting order parameter 
fully compatible with the observations. Considering its 
different structure from the known NCSCs, Zr$_3$Ir is expected to stimulate 
further studies on the interplay  
of space-, time-, and gauge 
symmetries in establishing the unique properties of NCSCs. 


This work was supported by the Schwei\-ze\-rische Na\-ti\-o\-nal\-fonds 
zur F\"{o}r\-de\-rung der Wis\-sen\-schaft\-lich\-en For\-schung (SNF) 
(Grants No.\ 20021-169455 and 206021-139082). 
S. K. G.\ and J. Q.\ are supported by EPSRC through the project ``Unconventional 
superconductors: New paradigms for new materials'' (Grant No.\ EP/P00749X/1).
L. J. C.\ thanks MOST for the funding under the projects 104-2112-M-006-010-MY3 
and 107-2112-M-006-020. Finally, we acknowledge the assistance from 
S$\mu$S beamline scientists at PSI.
\bibliography{ZrIr_bib}

\end{document}